\documentclass[letterpaper,twocolumn,10pt]{article}
\catcode`\D=12

\catcode`\D=11
\PassOptionsToPackage{breaklinks, colorlinks}{hyperref}

\usepackage[10pt,nocopyright]{sigmin}
\usepackage{libertine}
\usepackage{libertinust1math} %
\usepackage{inconsolata}

\usepackage[a-1b]{pdfx}
\usepackage{hyperref}
\usepackage{xcolor}
\definecolor{linkblue}{RGB}{0,40,200}
\hypersetup{%
  colorlinks=true,
  citecolor=linkblue,
  linkcolor=black,
  urlcolor=linkblue,
}

\input{glyphtounicode}
\usepackage[square,comma,numbers,sort&compress]{natbib}
\usepackage[T1]{fontenc}
\usepackage{xspace}
\usepackage{fancyvrb}
\usepackage{microtype}
\usepackage{graphicx}
\usepackage{amsmath,amssymb,amsthm}
\usepackage{aliascnt}
\usepackage{booktabs}
\usepackage[binary-units]{siunitx}

\usepackage{tikz}
\usepackage{tikzpeople}
\usetikzlibrary{shapes,arrows,automata,fit,calc,backgrounds,decorations.pathreplacing,calligraphy,patterns}
\usepackage{ifthen}

\usepackage{underscore}
\usepackage{relsize}
\usepackage{bytefield}
\usepackage{enumitem}
\usepackage{subfigure}
\usepackage{rotating}
\usepackage{chngcntr}

\usepackage{stfloats}
\usepackage{pifont}

\newcommand{\sys}{Shipwright\xspace}

\newcommand{\bool}{\mathsf{bool}}
\newcommand{\true}{\mathsf{true}}

\newif\ifdraft\drafttrue
\newif\ifnotes\notestrue
\ifdraft\else\notesfalse\fi

\definecolor{xxxcolor}{rgb}{0.8,0,0}
\definecolor{red}{RGB}{181, 23, 0}
\definecolor{blue}{RGB}{0, 118, 186}
\definecolor{gray}{RGB}{146, 146, 146}
\definecolor{green}{RGB}{26, 112, 30}
\makeatletter
\long\def\XXX{\@ifnextchar[{\@XXX}{\@XXX[]}}
\long\def\@XXX[#1]{\@ifnextchar[{\@@@XXX{#1}}{\@@XXX{#1}}}
\ifnotes
\long\def\@@XXX#1{{\color{xxxcolor} XXX #1}\xspace}
\long\def\@@@XXX#1[#2]{{\color{xxxcolor} XXX (#1) #2}\xspace}
\else
\long\def\@@XXX#1{\ignorespaces}
\long\def\@@@XXX#1[#2]{\ignorespaces}
\fi
\makeatother

\newcommand{\cc}[1]{\mbox{\smaller[0.5]\texttt{#1}}}

\fvset{fontsize=\footnotesize,xleftmargin=6pt}

\input{code/fmt}

\makeatletter
\newcommand{\oset}[3][0ex]{%
  \mathrel{\mathop{#3}\limits^{
    \vbox to#1{\kern-2\ex@
    \hbox{$\scriptstyle#2$}\vss}}}}
\makeatother

\makeatletter
\newcounter{@lena}
\newcommand{\lena}{\the\value{@lena}}

\makeatother

\theoremstyle{definition}

\newaliascnt{lemma}{thm}

\aliascntresetthe{lemma}

\newtheorem{theorem}{Theorem}

\def\Snospace~{\S{}}

\numberwithin{equation}{section}

\counterwithout{equation}{section}

  {\begin{list}{$\bullet$}%
     {\setlength{\parsep}{0pt}%
      \setlength{\topsep}{0pt}%
      \setlength{\itemsep}{0pt}}}%
  {\end{list}}

\bibpunct[: ]{[}{]}{,}{n}{XXX}{XXX}

\def\headline#1{\hbox to \hsize{\hrulefill\quad\lower0.5ex\hbox{#1}\quad\hrulefill}}

\renewcommand{\FancyVerbFormatLine}[1]{%
\ifnum\ifnum\value{FancyVerbLine}=101 1\else\ifnum\value{FancyVerbLine}=116 1\else0\fi\fi=1%
\raisebox{0.5em}{\headline{\textbf{\textrm{#1}}}}
\else#1\fi}

\RecustomVerbatimEnvironment
{Verbatim}{Verbatim}
{xleftmargin=2pt,baselinestretch=0.95}

\begin{document}
\title{\bf \sys{}: Proving liveness of distributed systems \\ with Byzantine participants}
\author{Derek Leung \\ MIT CSAIL \and Nickolai Zeldovich \\ MIT CSAIL \and Frans Kaashoek \\ MIT CSAIL}
\date{\vspace{-2\baselineskip}}

\maketitle

\begin{abstract}
Ensuring liveness in a decentralized system, such as PBFT, is critical,
because there may not be any single administrator that can restart the
system if it encounters a liveness bug.  At the same time, liveness is
challenging to achieve because any single participant could be malicious,
and yet the overall system must make forward progress.  While verification
is a promising approach for ensuring the absence of bugs, no prior work
has been able to verify liveness for an executable implementation of PBFT.

\sys{} is a verification framework for proving correctness and liveness
of distributed systems where some participants might be malicious.
\sys{} introduces three techniques that enable formal reasoning about
decentralized settings with malicious participants, allow developers to
decompose their system and proof in a modular fashion into sub-protocols
and sub-proofs, and support sound reasoning about cryptographic signatures that
may be embedded in messages.  We used \sys{} to implement and verify an
initial prototype of agreement on a single log entry in PBFT (with a few limitations)
and translate it to an executable implementation in Go.  We experimentally
demonstrate its operation and liveness both in the common case and
in several failure scenarios.

\end{abstract}

\section{Introduction}
\label{sec:intro}

Decentralized distributed systems operate in a challenging environment
where not all participants might behave correctly.  For example, Byzantine
fault tolerant protocols are designed to tolerate arbitrary misbehavior
from some fraction of participating nodes, such as in blockchains and
key-transparency systems.  In this setting, developers must ensure their
system achieves not just \emph{safety} but also \emph{liveness}---that
is, that the overall system keeps making progress.  Ensuring liveness
in the presence of Byzantine participants is hard, because malicious
participants can endlessly delay messages, retransmit the same messages
over and over, cause restarts, etc.  Moreover, many systems that worry
about Byzantine participants are \emph{open systems}, meaning there is
no single administrator in charge.  As a result, if a liveness bug were
to happen in a blockchain system or a key transparency system, there
would be no administrator that could restart all of the servers and get the
system running again.  Thus, for open distributed systems, liveness becomes a critical concern
at the same level as functional correctness (safety).

This paper introduces \sys{}, a new framework for machine-checked
verification of distributed systems, which enables developers to
prove both safety and liveness properties of their implementation.
Using \sys{}, we verify an implementation of the core
PBFT~\cite{castro:pbft-tocs} protocol.  The formal proof ensures that,
regardless of what the adversarial Byzantine participants do, the PBFT
system with nodes running our verified implementation will be correct
(safety) and will make progress (liveness) in all possible cases.

Consider an example mistake that we inadvertently made in our PBFT
implementation and caught with \sys{}.  When one view of PBFT
fails to reach consensus, a node can send a view-change message to other
participants in an attempt to make progress in the next view of
the protocol.  View-change messages in PBFT must prove that the sender
has seen enough messages from other participants to merit switching
to a new view.  This is done by including signed messages from other
participants as part of a view-change message.  Our implementation had a
bug where, in some situations, a node would include signatures from the
wrong previous view as part of a view-change message.  Other participants
would check the signatures in this view-change message, discover that
the signed message is not what they were expecting, and then reject it.
This would not lead to a safety violation---no invalid signatures would
have been accepted, and no incorrect outcome would be produced---but
it would lead to a liveness violation because the system might never
move on to the next view.

More precisely, the bug is triggered following a series of events.
First, the network drops all commit messages for some number of views
$k$ but permits prepare and view-change messages to propagate.
This prevents termination but causes some value $v$ to become
$\emph{prepared}$, which means it must show up in all subsequent view
changes.  Next, the network drops all prepare messages from view $k+1$.
Then, the network fully recovers for all messages in subsequent views
$k' > k+1$.  All nodes sign off on the view-change for value $v$
having prepared at
view $k$, which an honest primary attaches to the new-view message
kicking off view $k+2$---only for the nodes to reject the
signatures of the view-change messages (that they just created),
due to a bug in the byte-level
re-encoding of the view-change message plaintext within the new-view
message.  In fact, this holds for all future $k' > k+1$, so the
protocol never terminates.

As described in \autoref{sec:related},
\sys{} is the first framework that can verify liveness of distributed
system implementations and ensure the absence of bugs such as the one
described above.
\sys{}'s design addresses two key challenges.  First, \sys{} must
enable modular verification of liveness.  This means that developers can
use \sys{} to decompose overall protocols into smaller sub-protocols
that logically execute in parallel, such as the proposal, voting,
and view-change sub-protocols in PBFT~\cite{castro:pbft-tocs}.  This
reduces the amount of state that the developer has to reason about
at one time, enables proof automation, and allows reuse of components
(such as reusing the same broadcast primitive for both the prepare and
commit steps in PBFT).  Achieving this sort of modularity is a challenging
problem because liveness is a global property.  For example, classical
horizontal composition (combining two independent sub-protocols) does
not work well in the presence of a global clock that all sub-protocols
use for their timeouts, which in turn impacts liveness.  Similarly,
vertical layering of refinement proofs requires care in ensuring that
liveness properties are preserved, and, in particular, that there are no
infinite stuttering steps in a refinement proof.

Second, \sys{} must formally reason about Byzantine nodes.  To be sound,
\sys{}'s model must allow Byzantine participants to send arbitrary messages
over the network.  However, simply assuming that network messages are
untrustworthy is too strong because real systems rely on some fraction
of honest nodes to achieve safety and liveness.  Similarly, Byzantine
nodes cannot send an infinite number of messages that effectively stop
the system's progress by flooding the network.  Moreover, Byzantine
nodes cannot produce fake cryptographic signatures, which are often
necessary in these protocols for both safety and liveness; at the same
time, we do not want to explicitly model the details and probabilistic
definitions of signature scheme security, as this would substantially
complicate verification.

To address these challenges, \sys{} introduces several key techniques.
The first is an encoding of liveness conditions in terms of \emph{tasks}
that correspond to local \emph{completion measures}, which enables \sys{}
to translate liveness statements about infinite executions into proof
obligations about pairs of states.
These statements are more amenable to automated verification.
The second is a \emph{liveness-preserving composition operator}, which
enables developers to decompose complex protocols into simpler components,
such as composing the prepare and commit sub-protocols of a single PBFT
view, or composing an unbounded number of views to agree on a single
value in PBFT.
Finally, to soundly reason about Byzantine nodes, \sys{}
introduces a \emph{message stapling} model that captures how Byzantine
distributed systems use signatures in their messages without exposing
the low-level details of cryptographic constructions.

We implemented a prototype of \sys{} in Dafny.  To demonstrate the
benefits of \sys{}'s techniques, including composition and stapling,
our paper makes progress towards developing and proving a simplified
implementation of PBFT with \sys{}.
The implementation agrees on a single log entry, supporting an
infinite number of views, and its liveness guarantee ensures termination,
but it does not yet support agreement on multiple entries.  \sys{} proves
a top-level theorem that establishes a liveness-preserving refinement
between the executable implementation, operating on \sys{}'s network
model, and our abstract specification of PBFT.  Although PBFT is
our main case study, we believe \sys{}'s techniques
are generally applicable for reasoning about liveness of distributed
systems that use signatures.

We use Dafny's Go code extraction to generate an executable implementation.
Performance experiments shown in \autoref{sec:eval} demonstrate
that the implementation works but its performance is limited by
inefficiencies in Dafny's extracted Go code.  \sys{} consists of 13,357
lines of code, of which 1351 lines are trusted code.  The PBFT implementation
consists of 31,138 lines of code, which is largely proofs and internal
specifications; just 266 lines of code are trusted (e.g., the top-level
spec).

\section{Related work}
\label{sec:related}

The closest related work is the LiDO framework~\cite{qiu:lido}, which
takes a different approach to prove an end-to-end monolithic refinement
establishing the safety and liveness of the partially synchronous
Jolteon protocol~\cite{gelashvili:jolteon}.
LiDO is based on the \emph{ADO}~\cite{honore:ado} model and focuses on
consensus protocols,
while \sys{} is a general-purpose framework for systems---not
necessarily consensus protocols---with Byzantine faults.
LiDO proves liveness of Jolteon by decomposing its liveness properties
into the conjunction of safety properties and refinements of
\emph{timed} and \emph{segmented traces}. 
In contrast, \sys{} directly proves arbitrary weak fairness properties
by using completion measures.
The LiDO proof examines the implementation as a single piece of
code~\cite{lido-artifact}; \sys{} introduces a decomposition
operator that allows the implementation to be organized into
subprotocols and reused.
Finally, the LiDO proof models signatures at a high level but does not
implement them, as cryptography is absent from its
implementation~\cite{lido-artifact}, so it cannot reason about bugs
like the one described in \autoref{sec:intro}.
\sys{} enables a protocol developer to control and prove correct how
an implementation verifies signatures on messages' binary encodings.

IronFleet~\cite{hawblitzel:ironfleet} supports verification of some
liveness properties.
However, the systems verified in
IronFleet do not handle Byzantine faults and are relatively simpler in
terms of their liveness guarantees.
As a result, IronFleet was able to verify 
liveness with relatively simple and manual proofs built around
IronFleet's TLA encoding in Dafny. Furthermore, IronFleet's liveness
proof is separate from the correctness proof, and is not compositional
(i.e., the safety proof is a refinement, but the liveness proof is
not), which makes it difficult to prove liveness for applications
built on top of IronFleet's library. In contrast, \sys{}
handles PBFT, which is a more complex protocol and implementation,
requiring more sophisticated verification techniques to reason about
liveness in a Byzantine setting.

Bythos~\citep{zhao:bythos} is a Coq framework that takes a compositional
approach to liveness reasoning using local message passing and can
handle weak fairness properties.  Instead of verifying a partially
synchronous protocol (like PBFT), it verifies asynchronous protocols
which do not involve reasoning about time or timeouts in the code.
Bythos is able to leverage composition to prove end-to-end properties
about composite protocols, but the top-level theorems do not involve
refinement.  Bythos can generate executable implementations by extracting
Coq definitions of a protocol.

\paragraph{Protocol reasoning.}

There is a substantial amount of prior work on verifying liveness
properties of \emph{protocols} as opposed to \emph{implementations}.

Lamport presented machine-checked TLA proofs of safety refinement for
Byzantine Paxos, but did not construct machine-checked liveness
proofs~\cite{lamport:byzpaxos}.  \citet{bertrand:redbelly-verif}
verified safety and liveness of the Redbelly protocol using ByMC~\citep{konnov:bymc}, but did not
verify the implementation, and did not prove a succinct liveness specification;
instead, the top-level specification is a complex temporal-logic
statement, with a non-machine-checked argument of how this
temporal-logic statement relates to a simpler top-level specification.
Furthermore, \citet{bertrand:redbelly-verif} have a formal gap between
their partial-synchrony network assumptions and their fairness
assumptions, with a non-machine-checked argument for why they are
comparable.

\citet{losa:stellar} prove safety and liveness of the Stellar protocol, a
Byzantine-fault tolerant consensus protocol.  A key challenge they
address is to account for the capabilities of Byzantine nodes as
axioms in the decidable fragment of logic supported by the
Ivy~\citep{mcmillan:ivy, padon:ivylive} verification tool in order to
achieve full automation.
This work models Byzantine capabilities in terms of assumptions about how
they may affect non-malicious nodes' views of quorums.
\citet{berkovits:decomp} implement a tool that builds on Ivy to
automatically verify liveness properties for several Byzantine
consensus protocols, where the model of Byzantine behavior is captured
through similarly abstract axioms about quorums.  Verification in \sys
is less automated than the above. However, \sys targets
implementations instead of protocols, and \sys uses a lower-level
model of Byzantine faults in which Byzantine agents can craft
arbitrary messages (but not forge signatures).

\section{Design}
\label{sec:design}

This section presents the design of \sys{} by describing how it
handles three different levels: executable code, specifications, and
proofs.  At each level, we highlight how \sys{} supports three key
aspects: liveness, composition, and signatures.

\begin{figure}[ht]
  \centering
  \includegraphics[width=\linewidth]{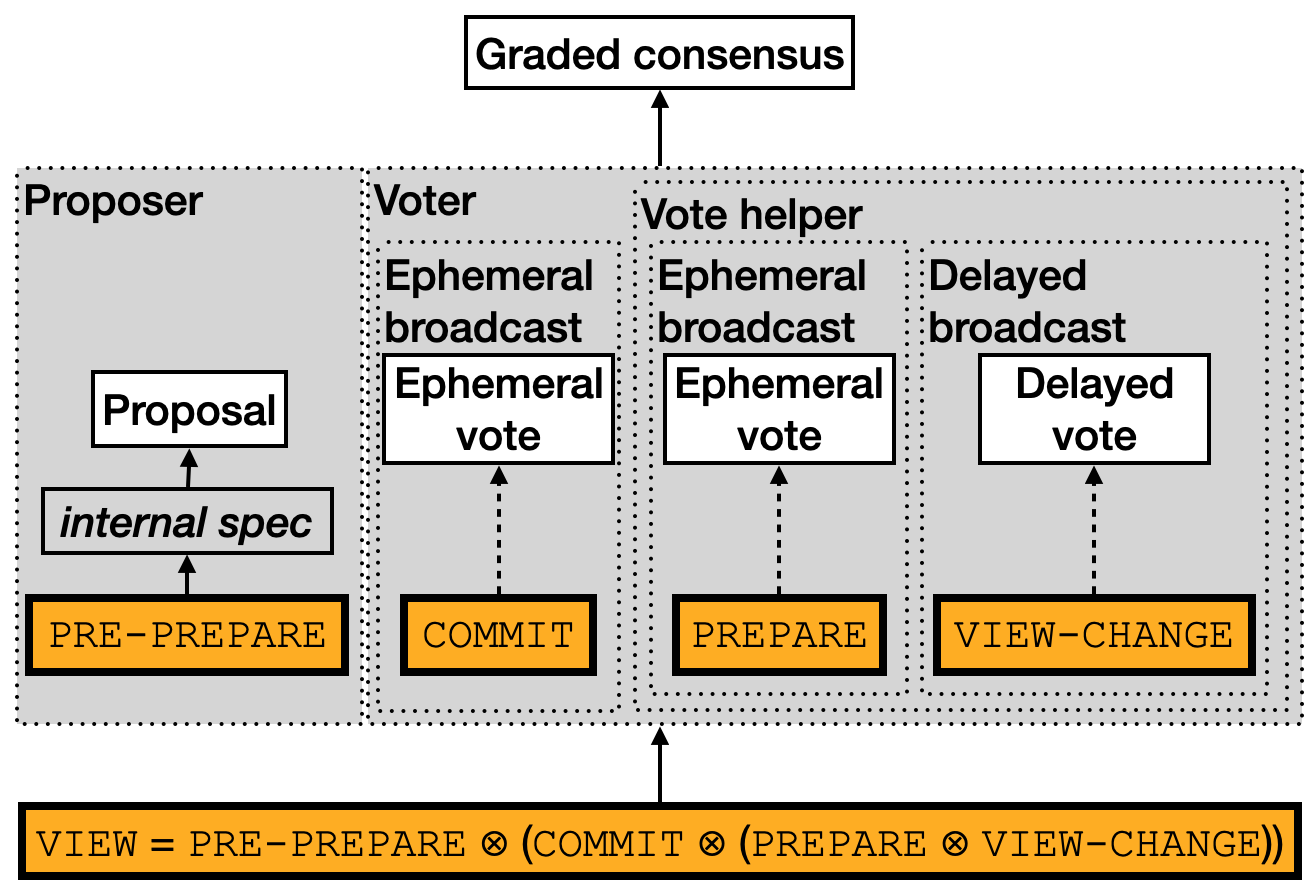}
  \caption{Modules in \sys as they are used to verify a single view of PBFT.
    $\otimes$ is \sys{}'s composition operator.
    (Not shown is how an infinite sequence of views compose
    into a single run of PBFT.)
    There is one module for each message type in PBFT,
    and the implementation of the entire view is constructed with
    synchronous dispatch into the implementations of
    each module.
    Only the internal structure of the top-level module and
    the \textsc{pre-prepare}-to-``proposal'' submodule are
    shown; all others are encapsulated.
  }
  \label{fig:modules}
\end{figure}

\sys{}'s overall goal is to allow a programmer to prove that a
complex system implementation can be thought of in terms of a simpler
abstract specification using \emph{refinement}~(\autoref{sec:spec-semantics}).
\autoref{fig:modules}
illustrates the intended use of \sys{}.  The developer writes
a modular implementation of PBFT, consisting of four sub-protocols
that comprise a single view (\textsc{pre-prepare}, \textsc{prepare},
\textsc{commit}, and \textsc{view-change}), and an unbounded sequence
of views that eventually agrees on a single value.  The developer then
writes specifications for each of these sub-protocols, and proves that
the implementations refine these specifications.  Finally, the developer
writes a top-level specification for the overall PBFT protocol, and
proves that the composition of the sub-protocol specifications refines
the top-level spec.  This proves that the low-level implementation is
a refinement of the top-level spec in terms of both safety and liveness.

\subsection{Executable code}

To implement a system, \sys programmers write three (pure) executable
Dafny \texttt{function}s, expressing their implementation in an
event-handling callback style, as shown in \autoref{fig:api}: a
node \emph{initialization} function, of type \cc{Zero}, which takes
some initialization parameters (such as node information and global
configuration); a node \emph{update callback}, of type \cc{Run}; and a
\emph{message authentication decoder}, of type \cc{StapledExtractor}.  Dafny tools
allow us to extract the \cc{Zero}, \cc{Run}, and \cc{StapledExtractor} functions
into a target language (Go in our prototype) and link them with a \sys{}
runtime that implements networking, signatures, timeouts, etc.

\begin{figure}[ht]
\input{code/api.dfy}
\caption{API for defining an implementation in \sys{}.}
\label{fig:api}
\end{figure}

The runtime calls the node update callback sequentially to handle
three kinds of events: (1) timeouts, (2) remote messages (received over
the network), and (3) local messages (received from a local program).
The update callback returns the new node state and a list of
messages, which are byte blobs, to send to peers.  The runtime signs
messages before transmission and verifies signatures before passing a
message to the update callback on the receiving node.

The runtime is written in Go and is not verified; it is assumed
to be correct.  It implements cryptographic primitives---both
signing messages and verifying signatures---and keeps track of the
node's cryptographic private key.  The runtime is also configured with
the identities of all peers---that is, a map from a \cc{NodeID} to an IP
address and a cryptographic public key.  We assume that this configuration
is common knowledge; it could be distributed via a trusted networking
and public key infrastructure.  \sys{}'s model assumes that all honest
nodes run \sys{}'s runtime with a verified implementation on top of it.
Byzantine nodes can execute arbitrary code but are assumed to lack
access to the private keys of honest nodes.

\paragraph{Liveness.}
To ensure liveness of executable code, \sys{} requires developers to
implement their code as \emph{total} Dafny functions.  This requires
the function to provably terminate on all inputs, ensuring that each
individual step (such as initialization or handling of some event)
will complete.

\paragraph{Composition.}
For the developers to reason about the correctness
of implementations of individual sub-protocols, \sys{} requires
that the implementation be clearly composed out of its constituent
sub-protocols.  To do this, \sys{} enables the developers to combine several
implementations of sub-protocols, each satisfying the executable API shown
in \autoref{fig:api}, into a single implementation that
again satisfies the same executable API from \autoref{fig:api}.

Composing protocol implementations poses two particular challenges.
First, the sub-protocols might not be entirely independent: for example,
when PBFT receives a quorum of votes in the \textsc{prepare} sub-protocol,
the execution model must allow the \textsc{commit} sub-protocol 
to learn about this fact.
Second, there may be an unbounded number
of sub-protocols being composed, such as an infinite number of possible views
in PBFT.  The execution model must allow representing this without materializing
an unbounded amount of state.

\sys{} addresses these challenges in its executable composition operator
with two ideas: \emph{synchronous dispatch} and a \emph{default map}
model.  The composition operator requires the developer to supply a
\cc{Tag} that names the sub-protocols being composed.  For instance,
composing two sub-protocols requires supplying a \cc{Tag} with two
possible values (e.g., \emph{left} and \emph{right}), whereas composing
an infinite number of view sub-protocols in PBFT would require supplying
a \cc{Tag} type of \cc{nat}, representing the view number.  The developer
also supplies the protocol implementation for each \cc{Tag}.

To efficiently handle composition of an unbounded number of sub-protocols,
\sys{}'s composition operator requires the developer to explicitly specify
the starting state of each sub-protocol in terms of a \cc{Zero} function.
This allows \sys{} to represent the overall state using a default map,
where some number of tags correspond to specific sub-protocol states
(e.g., views that have already started executing), but all other tags
(e.g., views that have not been reached yet) correspond to their default
zero values.

Logically, \sys{} delivers timeout events to all tags (sub-protocols)
in a default map, regardless of whether their state is zero.
However, for efficiency, it is important to avoid materializing state
for a tag in a map unless strictly necessary.  Thus, the developer must
prove that, in their implementation, a sub-protocol in a zero state takes
no action in response to timeout events (i.e., it transmits no messages
and its state remains zero).  This allows \sys{}'s runtime to
deliver timeout events only to tags explicitly present in the map.

To allow sub-protocols to interact with one another, \sys{}'s execution
model for a composed implementation---synchronous dispatch---
allows the developer to inject additional calls to sub-protocols when
handling inputs.  Specifically, \sys{} passes \cc{Timeout} inputs
to every sub-protocol and passes \cc{Message} inputs to the specific
sub-protocol targeted by the message (the composed implementation handles
messages of type \cc{(Tag, Message)} to disambiguate messages between
sub-protocols).  The developer can specify a function that, for any given
input, injects additional \cc{Call} inputs to any subset of \cc{Tag}s,
and those \cc{Call} inputs will be synchronously executed along with
the triggering input.  For example, the developer might specify that
a \cc{Call} to the \textsc{commit} sub-protocol be injected when the
\textsc{prepare} sub-protocol receives a vote that completes a quorum.

\sys{}'s composition operator attaches tags to messages, as described
above. It also adds tags when signing messages, which ensures that
transmitted and signed messages cannot be confused between sub-protocols.
For example, even though the \textsc{prepare} and \textsc{commit}
sub-protocols have the same executable implementation (they simply count
votes until a quorum is reached), their network messages and signatures
are distinguished by their \cc{Tag} value.

\paragraph{Signatures.}
\sys's model for reasoning about signatures must address several
constraints that arise in the context of proving liveness of
protocols like PBFT.  First, \sys must allow a node to explicitly relay
signatures from other nodes in messages that it sends.  This allows the
application to implement cryptographic certificates: for example, PBFT's
\textsc{view-change} message includes signed \textsc{prepare} messages from a
quorum of nodes to justify advancing the view.  We say these signatures
of \textsc{prepare} messages are \emph{stapled} to the \textsc{view-change}
message.  Second, \sys must allow the implementation to explicitly handle
signatures: storing them in node-local state, choosing which signatures to
send in a network message, etc.  Third, \sys should allow applications to
manipulate signatures separately from the messages that are being signed.
For instance, a certificate may consist of many signatures all signing
the same underlying message.  An application should not have to include
many copies of that message, or even any copy of that message at all,
if the recipient can infer the message from other context.  Moreover, we
do not wish to exactly model the correctness and security properties of
signing primitives themselves as they involve modeling complex assumptions
about probability and computational tractability~\cite{boneh:cryptobook}.
Finally, signature verification is computationally expensive and on the
critical path of the system, so we would like to enable its parallelism.

\sys{} addresses this challenge using \emph{message
authentication decoders}, which allow us to soundly reason about and
efficiently implement stapling.  A message authentication decoder is
an executable function with the type signature \texttt{StapledExtractor<Message,
  Signature>} as shown in \autoref{fig:api}.
This decoder computes the set of messages that are stapled to a particular
message, with their respective signatures and signers.
The sequence returned by the decoder is the set of
stapled signatures that go along with this message.  In each sequence
element, the \texttt{NodeID} value describes the node ID of the message
signer, the \texttt{Message} is the message whose signature is being
stapled, and the \texttt{Signature} is that signature.  \texttt{Message}
is a type parameter instantiated with the concrete message type
of the implementation to be verified, while \texttt{Signature} is left
opaque to the implementation.

The opacity of \texttt{Signature} allows the implementation to store and
produce \texttt{Message}s in an arbitrary way.  In particular, 
\autoref{fig:api} shows the update
callback type signature, \texttt{Run<Node, Message, Argument, Signature>},
where \texttt{Node} contains the node's
state, \texttt{Message} is the type of network messages, \texttt{Argument}
describes local \cc{Call} events, and \texttt{Signature} describes the
signature scheme.  The \texttt{Receive} and \texttt{Transmit} types,
also shown in \autoref{fig:api}, describe inputs to the callback and
output messages to transmit to peers, respectively.  When the callback
is instantiated with a particular implementation, \texttt{Node},
\texttt{Message}, and \texttt{Argument} can all depend on the opaque
\texttt{Signature} type parameter.  For instance, PBFT's view-change
messages can include stapled signatures (for a quorum of prepare messages).
To represent this, the \texttt{Message} type of the view-change
sub-protocol includes the sequence of stapled \texttt{Signature} values.

After verifying a root message's signature on receipt, the \sys runtime
calls the authentication decoder on the root message to obtain the set
of stapled messages and verifies their signatures before passing the
authenticated message to the update callback.

\subsection{Specifications}
\label{sec:approach:spec}

Specifications in \sys{} describe the allowed executions of an abstract
state machine, as shown in \autoref{fig:spec}.  The specification is
parametrized by three types. \cc{State} defines the state of a node.
\cc{Transition} (sometimes abbreviated \cc{Trans}) describes non-deterministic 
events that cause the state machine to handle an incoming network message
or a timeout, process a call from another sub-protocol, or observe
logical actions (such as the clock advancing or the network synchronizing).
\cc{Task} defines events that have liveness (fairness) properties
associated with them.

The \cc{Init}, \cc{Next}, and \cc{Invar} functions express safety properties,
defining the valid state-machine transitions for a specification.
\cc{Fair} defines the liveness specification, which says that each task
defined by the \cc{Task} type will eventually happen.

\begin{figure}[ht]
\input{code/spec.dfy}
\caption{A specification in \sys{} must consist of the four
  components shown in this code snippet.}
\label{fig:spec}
\end{figure}

To reason about the behavior of executable code, \sys{} provides a
specification for its low-level runtime, whose job is to invoke the
event handler functions of an executable implementation.  This boils
down to specifying the network, which involves the following:
digital signatures are unforgeable, real time eventually advances, all
nodes eventually receive timeouts, and all messages are delivered by a
deadline after stabilization.

\autoref{fig:net-spec} shows a fragment of our network specification.
The \cc{Init} and \cc{Next} functions say that the network runtime invoked
the implementation's \cc{Zero} and \cc{Run} functions, respectively,
or that an adversary injected a network message with \cc{Fault}.
\cc{Transition} includes \cc{Receive} and \cc{Transmit} events from
the executable implementation, as well as logical transitions such as
network stabilization (described below) and clock ticks.

\cc{Fair} captures \sys{}'s core liveness
assumptions.  In particular, \sys{} assumes a partial synchrony model
where the network will eventually stabilize, at which point all messages and timeouts
will be delivered by at most the network delay.  With that model in mind, \cc{Fair} in the
network specification says that the network will eventually stabilize,
that all messages and every timeout will be eventually delivered,
and that all clock
ticks will eventually happen.  For instance, the \cc{MessageF(m)} task
represents message \cc{m} being delivered; the \cc{FairMessage(st, tr,
mevt)} predicate states that the
message event \cc{mevt} (which includes some message \cc{m} as well as
its metadata) must be in the set of sent messages in state \cc{st}, but
not yet in the set of delivered messages, and that the transition \cc{tr}
must be the delivery of \cc{mevt}.

The network specification's \cc{Next} function also captures \sys{}'s assumptions about
signatures (not shown in
\autoref{fig:net-spec}).  Namely, \sys{} states that all messages sent over the
network have been signed with the sender's key.  This captures the fact
that the runtime will sign messages on transmission, and check signatures
on receive.  This also allows the developer to reason about signatures by
reasoning about the set of messages sent by honest nodes.

\sys{}'s network specification allows developers to reason about the
network in terms of arbitrary \cc{Message} types, but the lowest-level
network implemented by the \sys{} runtime requires that messages be
sequences of bytes---that is, \cc{Message = seq<bv8>} in Dafny.  Developers
can use a verified encoding/decoding library to show a refinement between
the network operating on byte-sequence messages and an abstracted network
operating on higher-level message types defined in Dafny.

\begin{figure}[ht]
\input{code/net-spec.dfy}
\caption{Fragment of a specification for the low-level network,
  including its liveness properties.
  We omit the details of \texttt{Next}, \texttt{Init}, and
  \texttt{Invar}.}
\label{fig:net-spec}
\end{figure}

\paragraph{Semantics of a specification.}
\label{sec:spec-semantics}

\sys defines an \emph{execution} to be an infinite sequence of
a system's \emph{states} $S$ and \emph{transitions} $T$, denoted
$((s_n, t_n))_{n \in \mathbb{N}} = ((s_0, t_0), (s_1, t_1), \ldots)$
where all $s_i \in S$ and $t_i \in T$.
A \emph{system specification} $\mathsf{Spec}$ captures a set of
executions of interest.
The syntax consists of a \emph{task set} $F$ along with four functions
$\mathsf{Init}: S \rightarrow \bool$,
$\mathsf{Next}: S \times T \rightarrow S$,
$\mathsf{Inv}: S \times T \rightarrow \bool$, and
$\mathsf{Fair}: F \times S \times T \rightarrow \bool$.

An execution \emph{conforms} to a specification when
(1) $\mathsf{Init}(s_0)$ holds;
(2) for all $i$, $\mathsf{Next}(s_i, t_i) = s_{i+1}$ and
(3) $\mathsf{Inv}(s_i, t_i)$ hold; and
(4) for all $f, i$, there exists a $j \geq i$ where $\mathsf{Fair}(f, s_j, t_j)$.

A \emph{refinement function} $r_{C,A}$ maps some set of states and
transitions to another set of states and transitions
$r_{C,A}: S_C \times T_C \rightarrow S_A \times T_A$.
We say that some \emph{concrete} specification \emph{refines} some
\emph{abstract} specification if for all concrete executions which conform to
the concrete specification, applying the refinement function
elementwise on the execution produces an execution which satisfies the
abstract specification.

A system's \emph{trace} is the sequence of transitions $(t_n)$ and
ignores the states.
A \emph{modular} refinement $\overline{r_{C,A}}$ is a trace refinement
exported by modules that hides the result of the refinement for $S_A$,
leaving only the result for $T_A$.

\paragraph{Terminal specifications.}

It is important for the developer to audit the top-level specification
used in \sys{}'s refinement proof to ensure that the machine-checked
proof is meaningful because the specification captures how the
developer expects the system to behave.  However, the general class
of specifications presented above can be challenging to audit for
two reasons.  First, the invariant in a specification constrains
reachable states \emph{and} transitions, which can be problematic if,
for example, an invariant says that certain messages cannot be delivered
in a particular state.  Second, the general fairness condition allowed
by our specification can, in principle, imply safety constraints, so
that the safety of a system depends on its liveness, which again makes
it difficult to audit the specification.

\sys{} addresses this challenge by introducing the notion
of a \emph{terminal} specification, taking inspiration from
Lamport's \emph{machine-closed} specifications of \emph{weak
fairness}~\cite{lamport:tla}.  Terminal specifications
must have a trivially true invariant and must only specify weak fairness
properties.  The developer supplies $\mathsf{Init}$ and $\mathsf{Next}$
as before, as well as a $\mathsf{WeakFair}$ in place of $\mathsf{Fair}$.
\sys{} then translates this to an ordinary specification as follows:
$\mathsf{Init}(s) = \mathsf{Init}(s)$,
$\mathsf{Next}(s, t) = \mathsf{Next}(s, t)$,
$\mathsf{Inv}(s, t) = \true$, and
\begin{align*}
\mathsf{Fair}(f, s, t) = (\mathsf{WeakFair}(f, s, t) \vee (\forall t, \neg (\mathsf{WeakFair}(f, s, t))).
\end{align*}
Lamport shows that such weakly-fair specifications
cannot imply additional safety constraints~\cite{lamport:tla}.

Note that \sys{} only requires terminal specifications only at the top
and bottom layers of refinement; internal refinement layers can use the
full generality of specifications, including invariants and arbitrary
fairness conditions.  This allows us to avoid quantifiers
in intermediate layers of the refinement proof, improving Dafny's
verification efficiency.

At the bottom level, \sys{} interprets executable code as a terminal
specification.  Executable code has no invariant, and the network
runtime associated with executable code only states weakly fair liveness
constraints, making the specification of executable code also terminal.

\paragraph{Composition.}

To enable the developer to reason about the composition of sub-protocols,
such as the four sub-protocols comprising a view of PBFT, or the
composition of PBFT's views, \sys{} provides a \emph{parallel composition}
operator, $\otimes$.  Parallel composition intuitively represents
executing two systems in parallel without any communication between them,
and the semantics are simply the concatenation of the semantics of the
components (so the new semantics are on $S_1 \times S_2$, $T_1 \times
T_2$, and $F_1 \cup F_2$).

\begin{figure}[ht]
\input{code/pbft-spec-join.dfy}

\caption{Event handling for composing an infinite sequence of PBFT views.}
\label{fig:composition}
\end{figure}

To allow the sub-protocols to interact, developers can interpose on
events to trigger additional calls on sub-protocols.  For example,
consider the composition of an infinite sequence of views in PBFT\@.
The developer needs to specify that, when one view completes, the next
view starts.  To do this, the developer supplies a \cc{Split}
function~(\autoref{fig:composition}) that determines what additional
\cc{Call} events must be delivered.  This \cc{Split} function captures
two important cases.  First, if the overall PBFT state machine receives
a \cc{Call}, it forwards it to view 0 to kick off the overall protocol
execution.  Second, if some view receives a message, and that message
happens to conclude that view, then \cc{Split} kicks off the next view.

\sys{} also allows the developer to strengthen a specification by
imposing an additional \emph{protocol invariant} on their transitions
$P: T_1 \times T_2 \rightarrow \bool$ (i.e., the invariant of
the new system requires all of $\mathsf{Inv}_1(s_{1_i}, t_{1_i})$,
$\mathsf{Inv}_2(s_{2_i}, t_{2_i})$, and $P(t_{1_i}, t_{2_i})$ to hold for
all $i$).  As before, we can consider an abstract protocol, between two
abstract systems, and the corresponding concrete protocol, between two
concrete systems.  Because \sys{}'s liveness conditions $\mathsf{Fair}$
are unconditional, parallel composition is purely syntactic: we do
not need to prove any additional verification conditions to establish
liveness when composing, even when adding a protocol invariant.

Recall that modules in \sys export not $r_{C,A}$ directly but rather a
trace refinement $\overline{r_{C,A}}$.  An advantage of exporting only
a trace refinement is it may be simpler than the state refinement:
the refinement only needs to be exposed to module clients only when
the abstract state changes, which may be much rarer than changes in the
concrete states.  Protocol invariants are also function of transitions
rather than states, which allows reasoning about them without considering
the details of the sub-protocol states.

For example, in order to reason about timely delivery of quorums,
the abstract object of the \textsc{prepare} module must maintain the
latest time any honest sender transmitted a \textsc{prepare} message.
This invariant is internally maintained by the module by computing
the maximum over all the transmission times of the \textsc{prepare}
messages on the network.  However, this abstract variable is set by
only one transition---the transition where the last honest node sends
the message---which only needs to inspect the current real time.

\paragraph{Signatures.}

\sys{}'s support for invariants in a specification, through the \cc{Invar}
predicate, allows the developer to abstractly reason about signed messages
and stapled signatures.  For example, in PBFT, a \textsc{view-change}
message requires the sender to include signed \textsc{prepare} messages
from a quorum of nodes to justify advancing the view.  \sys{}'s network
model allows the developer to capture this fact, by reasoning about
signatures and stapled messages.  However, reasoning at this low
level of abstraction---directly in terms of signed messages---can
be cumbersome.  Instead, a developer can use an invariant to state
that, if a \textsc{view-change} message is received, then a quorum
of other nodes must be in a state that agrees with this view change.
This frees the developer from explicitly reasoning about signatures
and allows reasoning directly about the state of other nodes and the
overall system.  To enable this reasoning, the developer will have to
prove that this invariant is implied by the low-level signatures, as
part of a refinement proof.

\subsection{Proofs}
\label{sec:approach:proof}

The top-level theorem we wish to prove is a refinement from the entire
implementation to a specification of a simple abstract object.
\autoref{fig:modules} illustrates the general proof strategy.
We start by refining the implementation into the parallel composition
of many modules, each implementing a subprotocol.
Next, for each module, we prove that the implementation refines a
corresponding abstract object.
Finally, we refine the parallel composition of the modules' now
abstract objects into our top-level simple abstract object.

\paragraph{Liveness.}

One particular challenge lies in proving facts about liveness in a way
that avoids reasoning about infinite executions.
Composition through protocol invariants allow us to combine smaller
theorems and thus arguments about liveness.
For individual specifications themselves, however, we
still need an effective means to reason about liveness.
In particular, our refinement theorems are about infinite executions,
which are hard for Dafny to reason about.
To optimize reasoning, we introduce reasoning principles which only
need to consider two states and transitions at once.

For safety properties, these verification conditions require showing the
compatibility of $r_{A,C}$ with the $\mathsf{Init}$, $\mathsf{Next}$,
and $\mathsf{Inv}$ functions, along with a \emph{refinement-local}
invariant $\mathsf{Inv_f}$.  In particular, it must be shown that
$\mathsf{Init}$ implies $\mathsf{Inv_f}$, and that $\mathsf{Next}$
preserves $\mathsf{Inv_f}$.  (This is similar to how many systems already
prove safety refinements, such as IronFleet.)

\begin{figure}[ht]
\input{code/measure.dfy}
\caption{Completion measure definition that implementations must satisfy.
  Developers must reason about one arbitrary state
  transition, \cc{trans} from state \cc{st}, and prove that for each
  liveness task \cc{task}, the measure \cc{mu} decreases to \cc{mu'}
  or the liveness property \cc{fair} is finally achieved.}
\label{fig:measure}
\end{figure}

A key novelty in \sys is that it supports a similar style of
verification conditions for \emph{liveness} in addition to safety.
Specifically, proving liveness in our new verification conditions
involves introduction of \emph{completion measures}, as shown in
\autoref{fig:measure}.
Every task in the task set is associated with a
\emph{completion measure} $m$, where $m$ is a member of a set $O_<$ equipped
with a well-founded partial order $<$.
At each state of the execution, the set of tasks is mapped to a set of
measures by the \emph{completion variant}
$\mathsf{Var}: F \times S \times T \rightarrow O_<$.
The programmer must then define a completion variant mapping
the concrete measure to an abstract measure with
the following property.

Consider any state $s_i$, transition $t_i$, and measures
$\mu_i, \mu_{i+1}$.
Suppose that, for all concrete tasks $f_C$, it is either the case that
$\mu_{i+1}(f_C, s_i, t_i) < \mu_{i}(f_C, s_{i+1}, t_{i+1})$ 
or $F_C(f_C, s_i, t_i)$ holds.
Then it must be the case that, for every abstract task
$f_A$, either
$\mathsf{Var}(r_{C,A}(s_i, t_i), f_A) < \mathsf{Var}(r_{C,A}(s_{i+1}, t_{i+1}), f_A)$
or $F_A(f_A, r_{C,A}(s_i, t_i))$ hold.

Because $\mathsf{Var}$ must be non-negative, this condition is strong
enough to imply liveness.
Moreover, as with safety properties, it mentions not an
infinite execution but rather two adjacent pairs of states and
transitions
$s_i, s_{i+1}, t_i, t_{i+1}$.

As an example, consider the voting sub-protocol of PBFT, which
is used for both \textsc{prepare} and \textsc{commit} messages.
\autoref{fig:vote-spec} shows a fragment of the specification for this
sub-protocol, defining three kinds of liveness tasks: stabilization of
the network (\cc{StabilizeF}), clock ticks (\cc{TickF}), and completion
of voting as seen by node \cc{id} (\cc{DoneF(id)}).

\begin{figure}[ht]
\input{code/vote-spec.dfy}

\caption{Fragment of specification for the vote sub-protocol in PBFT,
  focusing on its liveness.}
\label{fig:vote-spec}
\end{figure}

To prove that this voting sub-protocol implementation always reaches
\cc{DoneF}, the developer can define a measure counting down towards
\cc{DoneF} being reached.  In this example, the measure can be a
tuple $(|H(s)|, \mu(\mathrm{vote}(i)))$,
where $H(s)$ is the set of honest participants which this node
has not heard from yet,
$i$ is picked arbitrarily but deterministically from $H(s)$,
and $\mathrm{vote}(i)$ is the task that
completes when the node receives a \textsc{vote} message from node $i$.
\autoref{fig:vote-measure} shows the Dafny code defining this measure.
The measures for the \cc{StabilizeF} and \cc{TickF} tasks are
inherited from the network, but the measure for the \cc{DoneF} task
is a tuple consisting of the count of not-yet-voted nodes \cc{pending}
and the measure until we receive a vote from one of the pending nodes.

\begin{figure}[ht]
\input{code/vote-measure.dfy}

\caption{Measure used to prove liveness of voting sub-protocol
  in \autoref{fig:vote-spec}.}
\label{fig:vote-measure}
\end{figure}

After defining this measure, the developer simply proves that, for every
task \cc{task}, every transition \cc{t} from some state \cc{st} that
satisfies the invariant \cc{Invar(st, t)} will either decrease the measure
for \cc{task} or achieve the liveness condition required by \cc{Fair(task,
st, t)}.  The key benefit of the task-based completion measure approach
is that the \sys{} framework proves that this condition is sufficient to
establish liveness of all tasks in all possible execution traces, without
forcing the developer to reason about infinite traces.

\paragraph{Composition.}

A key benefit of \sys{}'s composition operators for code
and specifications is that \sys{} provides generic refinement
composition theorems.  In particular, if the developer has executable
implementations of sub-protocols $C_1$ and $C_2$, and proves refinement
of those implementations to abstract specifications $A_1$ and $A_2$
respectively, then \sys{}'s composition theorem proves that the executable
(synchronous-dispatch) composition $C_1 \otimes C_2$ refines the parallel
composition of the specifications $A_1 \otimes A_2$.  Similarly, \sys{}
provides a composition theorem for specifications, such that if $A_1$
refines $B_1$ and $A_2$ refines $B_2$, then $A_1\otimes B_1$ refines
$B_1\otimes B_2$.

\paragraph{Signatures.}

One key benefit of \sys{}'s signature model is that it allows \sys{} to
preserve simple reasoning principles for network messages---namely, if a
message is sent, the developer can soundly assume that the message will
be eventually processed by the recipient.  Ensuring this can be tricky
in the presence of stapled signatures in a message, because the receiving
node will check the stapled signatures and reject the message if any of
the signatures do not verify.  For instance, this was the issue behind
the bug described in \autoref{sec:intro}, where a node inadvertently
included signatures from the wrong view, causing the recipient to reject
the signatures as invalid.  To preserve the simple reasoning principles
about network message liveness, \sys{} requires that every transmitted
network message must have valid signatures for all stapled messages.
This helped us catch the aforementioned stapled signature liveness bug:
in the absence of this check at transmit time, the sender would have
been allowed to send messages with invalid stapled signatures, causing
the receiver to drop such messages and thus violate liveness.

\section{PBFT Case Study}
\label{sec:pbft}

To demonstrate \sys{}'s capacity for complexity,
we implement part of the PBFT protocol.
Specifically, the PBFT protocol establishes an ordered sequence of
commands in response to client requests to execute commands.
To do this, the protocol attempts to agree on each command in the
sequence.
If it succeeds, the command is appended to the sequence.
Otherwise, the special value \textsc{null} is appended.
\textsc{null} represents the absence of a valid command, and the
logical command sequence ignores \textsc{null} values as gaps.

We implement and verify a single iteration of this sub-protocol, which
we call \emph{broadcast}.
In \emph{broadcast}, a client sends a request to all servers.
The servers then execute a series of \emph{views} until
\emph{broadcast} returns a (possibly-\textsc{null}) value.

Each \emph{view} is an attempt to terminate the protocol
and is parameterized by a leader and a timeout.
\emph{Broadcast} terminates once (1) the network is stable, (2) the
timeout is sufficiently larger than the network delay, and (3) the
leader is honest.
Liveness thus hinges on the fact that (1) the
network always eventually stabilizes, (2) the timeout increases
exponentially, and (3) the leaders rotate in a cyclic order.

\subsection{Top-level theorem}

\begin{figure}[ht]
\input{code/pbft-spec.dfy}
\caption{Fragment of top-level specification for PBFT prototype.}
\label{fig:pbft-spec}
\end{figure}

As we describe in \autoref{sec:design}, our
\emph{concrete} state machine models the partially-synchronous
network instantiated with our PBFT implementation.
\autoref{fig:pbft-spec} describes our \emph{abstract} state machine,
which specifies correct behavior by
encoding three properties of interest:
\begin{itemize}
\item \emph{agreement}, the property that all honest nodes with a
  value agree on the same value (which is \texttt{st.reg});
\item \emph{termination}, the property that all honest nodes
  eventually terminate with some value (encoded by the task
  \texttt{TerminateF}); and
\item \emph{validity}, the property that if the client and first
  leader are honest, if the system is stable, and if a request was
  submitted early enough, then the system eventually agrees on the
  client's request (encoded with \texttt{st.send} and the task
  \texttt{SetF}).
\end{itemize}
(Note that if either the client or the first leader are malicious,
agreement holds but not validity, and this is represented by the
\texttt{Set} transition allowing an attacker's
value into \texttt{st.reg}.)

In order to connect the concrete state machine and the abstract state
machine, we need to prove a trace refinement theorem between the
concrete transitions and abstract transitions.
The trace refinement hides the network and code states, which are
quite complex.
Instead, the theorem exposes the mapping between the concrete and the
abstract transitions.
Among other things, this function maps a local call to an honest
client to the transition which sets \texttt{st.send}; it also maps the
receipt of a quorum of \textsc{commit} messages to the
\texttt{Terminate} transition.

We state the following top-level theorem in Dafny.
\begin{theorem}
  There exists a modular refinement $\overline{r_{C,A}}$
  between our PBFT implementation and our abstract specification.
\end{theorem}

\subsection{Decomposition strategy}

\begin{figure}[ht]
  \centering
  \includegraphics[width=0.6\linewidth]{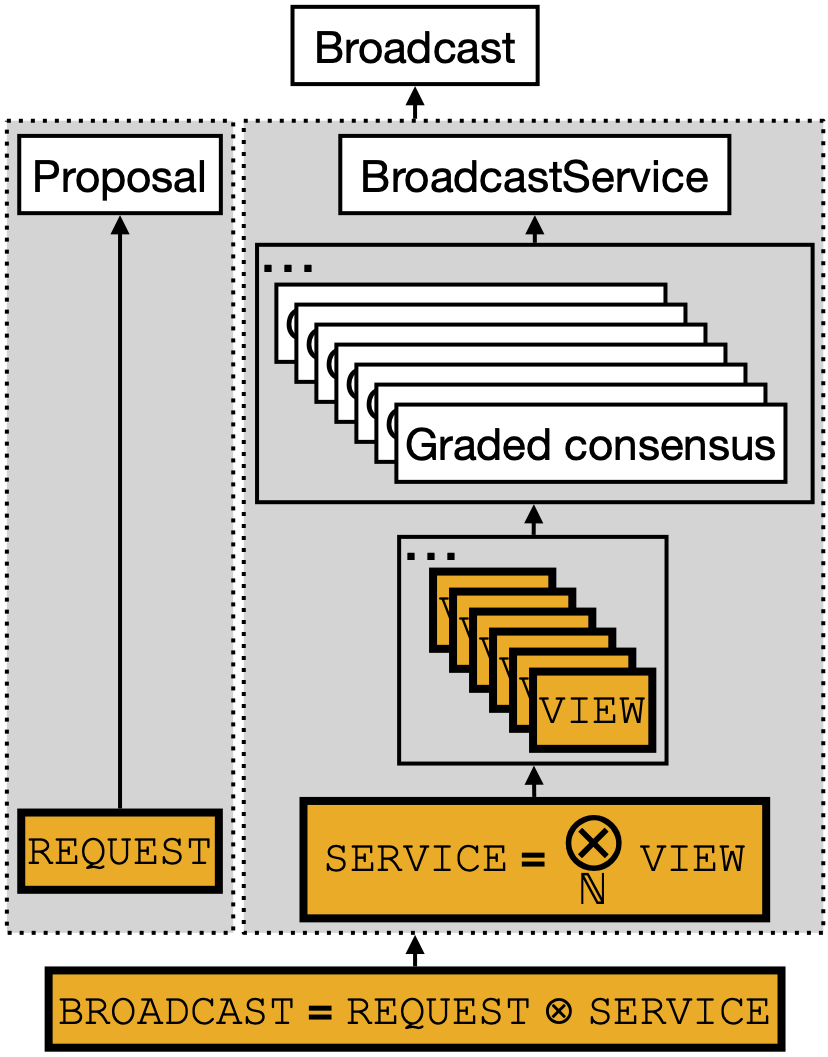}
  \caption{A diagram of the refinement proof's compositional structure;
    \autoref{fig:modules} shows the detailed structure of the refinement
    between a single view implementation and a single graded consensus
    spec.}
  \label{fig:toplevel-modules}
\end{figure}

In order to make our proof verification overhead tractable to Dafny, we decompose
the entire program as follows.
First, we decompose the client request from the rest of
\emph{broadcast}.
Second, we decompose the broadcast service to an infinite map of view
implementations.
Third, we decompose the views themselves into their constituent
messages as in \autoref{fig:modules}.
\autoref{fig:toplevel-modules} illustrates this strategy.

Our proof broadly follows the structure of the decomposition.
We create an abstraction for each individual message type and prove
that our sub-protocol for that message type refines into the
abstraction.
We are able to reuse modules where implementations are identical.
For instance, the \textsc{prepare} and \textsc{commit} sub-protocols
have identical implementations; so do the client \textsc{request} and
the \textsc{pre-prepare} sub-protocols.
For each message type, we prove liveness properties by constructing
measures which rely on the underlying liveness of message delivery
as well as by counting down the remaining messages yet to be received.
\autoref{fig:vote-spec} illustrates the measure for \textsc{prepare}
and \textsc{commit}.

Once we have atomic, per-message-type refinements, we combine them
together using \sys{}'s composition operators, which are liveness-preserving.
We build a view out of PBFT's message types.
Views guarantee that \textsc{view-change} quorums are always
eventually delivered once all honest nodes have started the view.
In addition, if the validity conditions hold, then the view guarantees
that \textsc{commit} quorums are also always eventually delivered.

We build the broadcast service out of an infinite sequence of views.
The invariants on the state machine transitions ensure that whenever a
quorum of \textsc{view-change} messages are received in some view, the
timeouts for the following view immediately begin.
\autoref{fig:composition} illustrates the code which implements this.
This concrete constraint here is mapped to the corresponding abstract
constraint via the view's trace refinement theorem.
At the abstract level, it is then possible to prove that the infinite
sequence of views refine the broadcast service by incorporating into
the measure (1) the concrete measure representing stabilization, (2)
the difference between the current view timeout and the timeout of the
first view long enough to accommodate the network delay, and (3) the
difference between the current view and the first view with an honest
leader.
Before all three conditions are met, we rely on the liveness of
\textsc{view-change} quorums.
Once they are met, we proceed to terminate the protocol.

Finally, we show that coupling the client with the broadcast service
refines \emph{broadcast}.
Proving termination here requires composing the termination of the
abstract client proposal with the termination of the broadcast
service.

\paragraph{Limitations.}
There are two limitations in our current PBFT prototype, stemming
from the fact that our prototype implements just one iteration of PBFT
instead of an append-only log over multiple entries.
First, our top-level theorem does not currently capture the reply
messages sent to clients once a majority of honest nodes has
seen a \textsc{commit}-quorum.
(A client needs $f+1$ signed replies to conclude that the request was
successful.)
Second, in order to state and prove an end-to-end liveness property,
we need to allow for multiple iterations of \emph{broadcast} itself,
since a malicious initial leader can force the view to commit with
\textsc{null}.
We expect both of these limitations would be addressed by extending our
prototype to agree on multiple entries in PBFT.

\subsection{Bugs Found}

We found several bugs during the verification process.

As described in \autoref{sec:intro}, we discovered a liveness bug
where, when sending a \textsc{new-view} message, a node would attach
signatures from an incorrect previous view.
Attaching the incorrect signatures caused the receiver to reject
otherwise valid messages, preventing the start of the next view.

We caught another two bugs related to certificates in which we would
count signatures incorrectly.
The implementation accepted certificates with too few signatures,
which could allow a malicious node to convince an
honest node to accept any value.
We were able to catch this bug because our model of cryptography
does not assume that non-cryptographic message validation
is correct.

Another liveness bug relates to timekeeping.
In some cases, a node would start its timer too early
and thus time out of the next view prematurely.
No view would complete in time, even after the
system had stabilized.

Finally, we caught another bug also related to timeouts.
A node sends a \textsc{view-change} message when either the view times
out or when it receives $f$+1 votes.
Crucially, the node can only send a \textsc{view-change} message on
exactly one value.
An early implementation sent \textsc{view-change} messages on
\emph{both} of these events without checking that their values were
the same.
Depending on the implementation of the \textsc{commit} messages, this
may have resulted in a safety violation as a quorum of \textsc{null}
view-change votes could have coexisted with a quorum of
\textsc{commit} messages.

\section{\sys{} Implementation}
\label{sec:impl}

\begin{figure}[ht]
  \centering
  \begin{tabular}{@{}lrrr@{}}
    Component & Trusted & Untrusted & Total \\
    \midrule

    \bf{\sys{} framework} & \textbf{1351} & \textbf{12006} & \textbf{13357} \\
    \hspace{3mm}utilities & 9 & 842 & 851 \\
    \hspace{3mm}state machines & 650 & 2271 & 2921 \\
    \hspace{3mm}environment model & 309 & 0 & 309 \\
    \hspace{3mm}composition & 0 & 8893 & 8893 \\
    \hspace{3mm}Go runtime & 383 & 0 & 383 \\
    \midrule

    \textbf{PBFT case study} & \textbf{266} & \textbf{30872} & \textbf{31138} \\
    \hspace{3mm}top-level spec & 266 & 0 & 266 \\
    \hspace{3mm}sub-protocol specs & 0 & 10337 & 10337 \\
    \hspace{3mm}sub-protocol proofs & 0 & 18647 & 18647 \\
    \hspace{3mm}implementation & 0 & 1888 & 1888 \\
    \midrule

    \textbf{Total} & \textbf{1617} & \textbf{42878} & \textbf{44495}
  \end{tabular}

  \caption{The number of lines of code necessary to implement the
    trusted and untrusted portions of our implementation.
    Total figures are in bold.}
  \label{tbl:loc}
\end{figure}

\autoref{tbl:loc} summarizes the implementation of \sys{} and the PBFT
case study, and their breakdown in terms of  major components.  The \sys{}
framework consists of roughly 13300 lines of code, of which around 10\%
are trusted.  The trusted parts include the definition
of state machines and the network in Dafny, and the Go implementation of
the \sys{} runtime.  The PBFT case study consists of roughly 31200 lines
of code, of which just under 270 lines of code are trusted---namely,
the top-level specification.
The rest of the PBFT implementation and proofs are not trusted.
Aside from our code, we also trust the implementations of systems
components such as the Dafny and Go compilers, the Go libraries and
runtime, the OS kernel, the networking stack, the hardware, etc.

The \sys{} framework is divided into two main libraries.
The first library defines foundational concepts such as state
machines, executions, liveness, and trace refinement.
It includes reasoning principles, proven sound, that enable safety and
liveness reasoning with measures.
It also describes how to compose state machines with the $\otimes$
operator and how to strengthen protocol invariants.
This library describes concepts that are independent of particular
assumptions about the environment under which a system operates.

The second library, which depends on the first, specializes the
state machines to our environment.
This includes our model of partial synchrony and the security of
digital signatures.
It defines the API (\autoref{fig:api}) against which an application
developer must write code.

It also establishes how code can be soundly transformed under our
formalization of state machines.
The library contains code implementing synchronous dispatch
and specifies how dispatch functions correspond to
the protocol invariants between state machines.
At the lowest level, the library contains (untrusted) routines for
producing canonical byte encodings of network messages and a proof
that a corresponding refinement follows from these encodings.
It was useful to prove these refinements sound in a generic
manner---not only to ease proof reuse, but also so that we could hide
complex compositional definitions from the Dafny verifier in a
principled manner in order to speed up verification.

We provide a trusted implementation of the runtime in the Go
programming language.
This runtime implements our untrusted code's server environment.
It uses the standard libraries
\texttt{time}, \texttt{net}, and \texttt{crypto/ed25519}
to implement timing, networking, and cryptography, respectively.

In addition to the main components, we provide
library code for a variety of utilities for
transforming Dafny collections
such as sets, sequences, and infinite maps.
These libraries allow us to
manage quantifier instantiation, which has a
significant impact on Dafny verification performance.
We also implement a library specialized for efficient automated
reasoning about quorums.

\paragraph{Limitations.}
The proof of the refinements in the PBFT prototype to and from the
broadcast service are works-in-progress and are not yet complete.
We have not yet proven correct all of the trace properties required for
safety and liveness of the end-to-end specification.

\section{Evaluation}
\label{sec:eval}

Our evaluation attempts to answer the following questions.
\begin{itemize}
\item How does our implementation perform, end-to-end?
\item How does our implementation preserve liveness when there are
  attacks by participants or on the network?
\end{itemize}
We evaluate our PBFT implementation on a virtual machine with 24
cores, 6~GB of RAM, a Intel Xeon E5-2699 CPU, and running Ubuntu 22.04.
All of the PBFT nodes run on the same virtual machine and connect to
each other via localhost.  Each node is a Linux process, and the nodes
are fully connected to each other through all-to-all TCP connections.
Nodes send messages to themselves using an asynchronous Go goroutine.
Every 250ms, we deliver a timeout event to a node with the current time.
We set the initial PBFT view timeout to 1s.  We use Dafny version 4.9.1
and Go version 1.24.2.

\subsection{Common Case}

\begin{figure}[ht]
  \centering
  \includegraphics[width=0.8\linewidth]{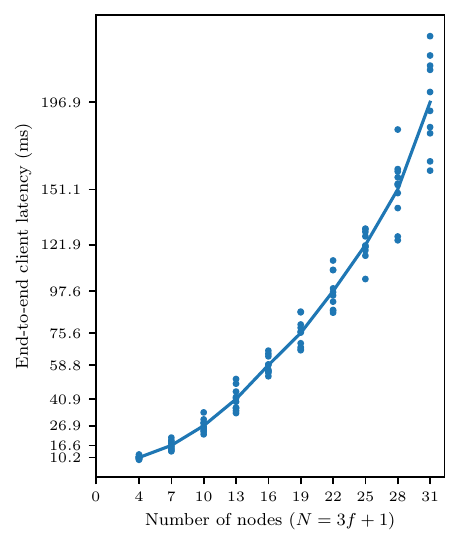}
  \caption{End-to-end latency of a client request in a system with
    $N = 3f+1$ nodes, varying values of $f$ from 1 to 10 for 10
    iterations each.
    The scatter plot represents the raw data while the line plot
    represents averages.
    The labels on the y-axis correspond to the averages.}
  \label{fig:eval-fastpath}
\end{figure}

To evaluate common-case performance of our PBFT prototype, we ran a
PBFT system consisting of $3f+1$ nodes, varying $f$ from 1 to 10.
When the system is started, one node immediately broadcasts a client
request to all other nodes.
\autoref{fig:eval-fastpath} reports the time (in ms) it takes for this
node to receive $f+1$ signed replies over 10 trials.

In this scenario, there are no failures, so the time includes five
message types: the client request, the \textsc{pre-prepare},
\textsc{prepare}, and \textsc{commit} messages, and the signed service
response.
Thus there are $1 + 2(2f+1) + 2(2f+1)^2$ total messages on the critical path
(counting messages from nodes to themselves).

\subsection{Performance Under Failure}

\begin{figure}[ht]
  \centering
  \includegraphics[width=\linewidth]{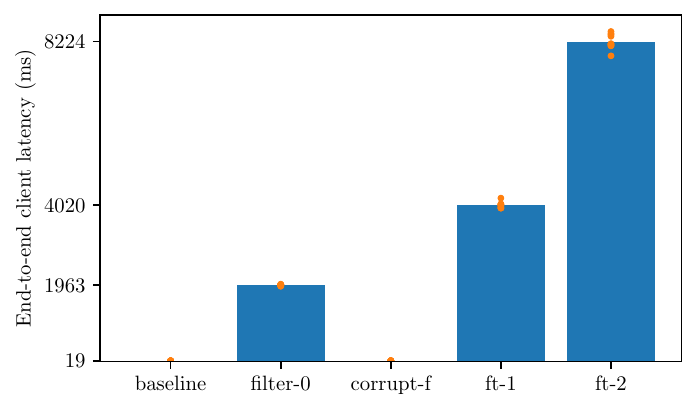}
  \caption{End-to-end latency of a client request under a variety of
    scenarios representing different attacks, given $f=2$.
    In the first scenario, an attacker intermittently prevents the
    delivery of \textsc{commit} messages in view 0, but the system
    recovers shortly afterwards.
    In the second, the attacker controls two nodes.
    In the last two scenarios, both the network attack and node
    corruption take place (for one and two nodes, respectively). }
  \label{fig:eval-failures}
\end{figure}

To measure the effects of failure on the system liveness and
end-to-end performance, we simulate three scenarios representing various
attacks on $3f + 1 = 7$ nodes (setting $f = 2$).
Since our implementation only covers agreement on a single value in the PBFT
protocol (as opposed to
many values forming an entire replicated log), we consider
scenarios where we will obtain a non-\textsc{null} value.

In the first case, we consider a network under some denial-of-service
attack.
In order to prevent the system from making progress, an attacker
forces the network to drop all \textsc{commit} messages from view 0.
The network recovers shortly afterwards.
We expect the system to time out, enter view 1, and for the view 1
leader to recover with the correct value.

In the second case, we suppose that $f$ of the nodes, which are neither
the client nor the view-0 leader, are malicious.
This node sends \textsc{pre-prepare} (and the attached
\textsc{new-view}), \textsc{prepare}, \textsc{commit}, and
\textsc{view-change} messages on the incorrect value.
Under this attack, we expect the system to commit the correct client
value in view 0.

In the third case, we consider both failures simultaneously.
We explore scenarios in which the first $f$ leader nodes after view 1
are malicious.
In all cases, we expect the system to recover in the first view with
an honest leader the malicious ones (i.e., views 2 and 3,
respectively).

\autoref{fig:eval-failures} illustrates our results.
As expected, our clients experience an end-to-end latency consistent
with the sizes of views.
In particular, when the network is faulty, our clients receive a
response at either 2s, 4s, or 8s, depending on how many nodes are
malicious.
Moreover, when the network is not faulty, end-to-end latency is
unaffected, as we expect.

\section{Conclusion}
\label{sec:concl}

We design and implement \sys{}, a framework for proving liveness of
distributed systems with Byzantine participants.
\sys{} allows developers to prove that implementations of complex
protocols, like PBFT, refine a simpler high-level specification, by
introducing tools to manage the burden of proof.
First, developers can decompose larger protocols into smaller ones
in a liveness-preserving manner.
Second, completion measures enable them to prove liveness properties
about whole executions via proof obligations over pairs of states.
Third, \sys{} allows developers to prove that their implementation
uses digital signatures correctly (e.g., in message stapling) without
resorting to low-level cryptographic details.
We use \sys{} to verify our executable implementation that agrees on a
single log entry in the PBFT protocol.

\bibliographystyle{abbrvnat}
\bibliography{n-str,p,n,n-conf}{}

\end{document}